# Wafer-scale fabrication and room-temperature experiments on graphene-based gates for quantum computation

Mircea Dragoman, Adrian Dinescu, and Daniela Dragoman

*Abstract*—We have fabricated at wafer scale graphene-based configurations suitable for implementing at room temperature one-qubit quantum gates and a modified Deutsch-Jozsa algorithm. Our measurements confirmed the (quasi-)ballistic nature of charge carrier propagation through both types of devices, which have dimensions smaller than the room-temperature mean-free-path in graphene. As such, both graphene-based configurations were found to be suitable for quantum computation. These results are encouraging for demonstrating a miniaturized, room-temperature quantum computer based on graphene.

## I. Introduction

QUANTUM computers are studied intensively by renowned computer companies and many research groups, and optimistic announcements about the eminence of progress from lab phase to small scale production are presented monthly in many journals and media [1]. However, the main feature of a classical computer, termed as von Neumann computer, namely miniaturization, will be lost if the future of quantum computers will be based on ion traps or superconducting technologies [2], the most advanced technologies to date for quantum computing.

Could we perform quantum computations via quantum gates based on solid-state technology at room temperature, with nanoscale devices, to preserve the unprecedented miniaturization of electron devices, which allows today the integration of billions of transistors on chip, on a surface of few $cm^2$ [3]? The answer is not straightforward, because the main adverse effects against quantum computation are the laws of quantum mechanics themselves. In particular, decoherence, i.e. the rapid loss of coherence of quantum states, is the most detrimental effect against quantum computing. Therefore, superconducting devices working at very low temperatures, with long decoherence times (of few milliseconds), are the most advanced implementation of quantum computing, thousands of gates based on Josephson junctions being able to perform quantum computations with multiple qubits. Although other solid-state technologies are presently considered for quantum computation, such as spintronics [4] based on different materials and especially on Si [5], the implemented quantum gates still work at very low temperatures. So, the paradigm of quantum computation resides in the computing time of few milliseconds, during which millions of operations take place, and after which decoherence installs itself and the quantum computer must be refreshed with new quantum coherent states.

A solution of these issues was proposed in Refs. [6] and [7], where we have shown that quantum gates and algorithms, such as the Deutsch-Jozsa quantum algorithm, could be implemented using ballistic transport in graphene. The ballistic or quantum transport is a transport regime where electrons behave as coherent quantum waves in the absence of scattering, until the mean-free-path is attained. The main advantage of the ballistic transport regime is that quantum wavefunctions do not decohere, so that quantum superposition is preserved throughout the structure. Actually, the coherence/quantum superposition of qubits in a quantum computer is preserved even if few scattering events take place, in the so-called quasi-ballistic regime, since the key parameter for coherence survival is the phase coherence length [8].

The room-temperature mean-free-path is generally very small, e.g. Si has a mean-free-path of only few nanometers at room temperature. However, there are semiconductors, such as InAs nanowires or InSb/AlInSb and InAlN/GaN heterostructures, where the mean-free-path is in the range of 100-500 nm at room temperature [9]. Graphene is by far the material with the largest mean-free-path at room temperature known today. The room-temperature ballistic transport regime in high-quality graphene monolayers is preserved for mean-free-paths up to 400 nm if CVD graphene is transferred on a $SiO_2$ substrate [10], or well beyond 1 µm in graphene encapsulated in boron nitride [11] or grown on SiC [12]. On the other hand, the phase coherent length is always larger than the mean-free-path, since coherence loss cannot occur without

Manuscript received XXXX; revised XXXXX; accepted XXXX. Date of publication XXXXX; date of current version XXXX. This work was supported by the Romanian Ministry of Research and Innovation, under Grant TEHNOSPEC PN 1632 2015-2017 (Nucleus Project).

M. Dragoman and A. Dinescu are with the National Institute for Research and Development in Microtechnology (IMT), 077190 Bucharest-Voluntari, Romania (e-mail: mircea.dragoman@imt.ro; adrian.dinescu@imt.ro).

D. Dragoman is with the Faculty of Physics, University of Bucharest, 077125 Bucharest-Magurele, Romania, and with the Academy of Romanian Scientists, Splaiul Independentei 54, 050094 Bucharest, Romania (e-mail: daniela@solid.fizica.unibuc.ro).

scattering [8], and thus a quantum device working in the ballistic regime, i.e. below the mean free path, will always work below the phase coherence length.

In this paper, we present the wafer-scale fabrication of and first measurements on quantum gates based on ballistic transport in graphene at room temperature, following the configurations proposed in Refs. [6] and [7]. In particular, we fabricated ballistic electron interference devices that could implement quantum gates (QG), such as one-qubit Hadamard or NOT gates depending on the length of the interference region, or a one-qubit modified Deutsch-Jozsa (DJ) quantum algorithm. The results presented in this paper demonstrate that the fabricated graphene-based configurations are suitable for quantum computation.

## II. WAFER-SCALE FABRICATION OF GRAPHENE QUANTUM GATES

We have fabricated 25 QG and 25 DJ quantum devices using a graphene chip cut from a 4 inch graphene wafer having as substrate a heavily doped Si layer over which 300 nm thick $SiO_2$ was thermally deposited. The graphene transfer on the 4 inch wafer was performed by Graphenea. The main technological steps involved in the fabrication of graphene quantum gates at the wafer scale are: 1) patterning of the alignment marks by e-beam lithography (EBL), 2) metal deposition (Ti/Au – 10 nm/90 nm) and lift-off, 3) graphene shaping by EBL and reactive ion etching (RIE), 4) contact pads patterning by EBL by depositing a thick (300 nm) PMMA layer on the substrate, exposing it to the e-beam and then developing it, and 5) metallization of 10 nm of Ti and 100 nm of Au by e-beam evaporation in a highly directional PVD machine (Temescal FC 2000), and lift-off.

Figure 1(a) shows the result of graphene patterning, process performed by covering the wafer with a 100-nm-thick layer of PMMA and a 40-nm-thick layer of HSQ, irradiating it by the e-beam at 30 kV and 200 uC/cm$^2$ in a dedicated EBL equipment (RAITH e_Line), followed by the development of the HSQ layer, etching away of the PMMA and graphene by RIE and, finally, by the lift-off process of the HSQ in order to uncover the patterned graphene. SEM image details of the patterned three-terminal QG and DJ devices are represented in the lower part of Fig. 1(a). Figures 1(b) and 1(c) show details of the structure after patterning of the contact pads (step (4) above) and, respectively, their metallization (step (5) above), whereas Fig. 1(d) illustrates the entire three-terminal devices, with one input port denoted by *in* and two output ports labelled as *out*1 and *out*2.

The relevant dimensions of QG are: the length of the entire device is about 340 nm, with the interference region (IR) of about 180 nm. In the case of DJ, the entire device is about 370 nm long, the IR length being 200 nm. Thus, the total length of both quantum devices are less than the mean-free-path of graphene at room temperature, i.e. 400-500 nm [13]. The widths of the Y-junction branches, which form electron waveguides/nanoribbons, are 100 nm and 140 nm, respectively, for QG and DJ devices.

In both QG and DJ devices, the input Y-junction, *in*,

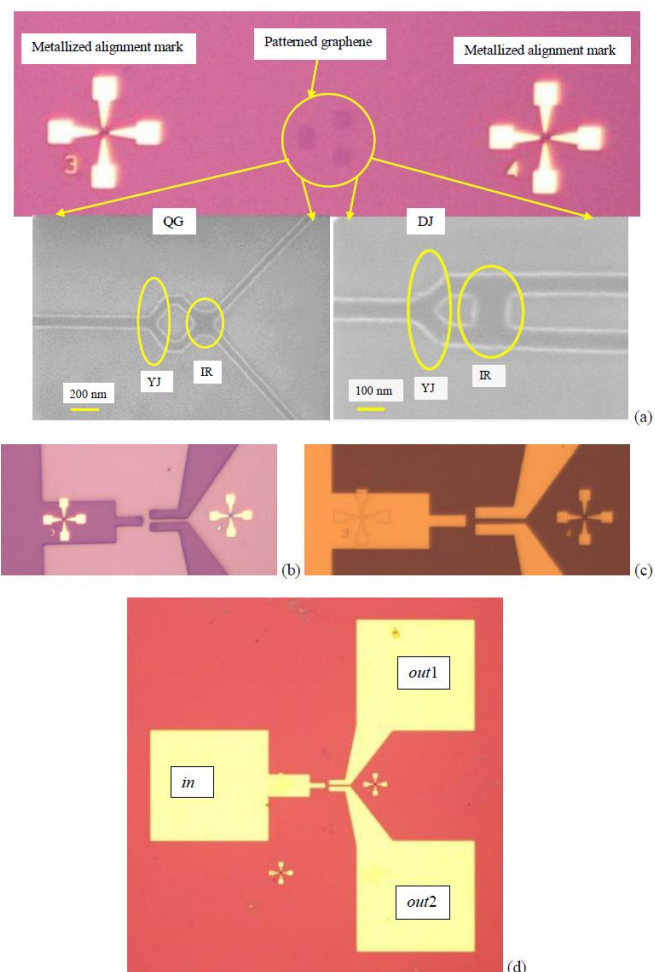

Fig. 1. Illustration of some fabrication steps of quantum graphene devices: a) graphene patterning of QG or DJ devices, (b) patterning and (c) metallization of the contact pads, and (d) the final three-terminal device.

denoted by YJ in Fig. 1(a), creates a superposition of $|0\rangle$ and $|1\rangle$ quantum logic states, identified as the upper and lower branches of the Y-junction.

More precisely, the quantum states/qubits $|0\rangle$ and $|1\rangle$ are the quantum wavefunctions in the upper and lower branches, which result from the incident wavefunction by splitting in the Y-shaped graphene monolayer in the ballistic transport regime of carriers. The two branches must allow the propagation of only one mode. This requirement can be fulfilled by appropriately choosing the width $w$ of the Y-junction branches and/or the applied potential on the structure, $U_G$, induced by a backgate voltage that covers the entire structure, since the number of propagating modes for electrons with energy $E$ is given by the integer part of $(E-U_G)w/(\pi\hbar v_F)$, as shown in [7]. Here, $v_F = 10^6$ m/s is the Fermi velocity of charge carriers in graphene. Note that the simulations in Refs. [6] and [7] of charge carrier transport in Y-junctions were performed using the Dirac Hamiltonian for graphene in the continuum model, while the same configuration could be modeled also by a scattering matrix approach in an optimized discrete formulation [14].

A symmetric junction implements the $(|0\rangle+|1\rangle)/\sqrt{2}$ input quantum state, any geometric or electric-field-induced asymmetry in the splitting region generating a different superposition. The quantum logic states $|0\rangle$ and $|1\rangle$ enter the wider interference region, denoted by IR, the resulting wavefunction being split by an output Y-junction into two parts that emerge from the *out*1 and *out*2 branches. Although the QG and DJ devices look similar there are two main differences between them, implying different requirements in the design:

*r*1) a meaningful QG device of Hadamard or NOT type requires an input electron wavefunction incident on only one arm of the Y-junction, i.e. an initial $|0\rangle$ or $|1\rangle$ state, i.e. it needs an asymmetric Y-junction, or an in-plane electric field transverse to the input waveguide, which could tune the superposition of the $|0\rangle$ or $|1\rangle$ state, whereas a DJ device requires a symmetric Y junction, i.e. an initial $(|0\rangle+|1\rangle)/\sqrt{2}$ state.

*r*2) the interference region of the QG device is designed to support only two modes, whereas that of the DJ device can be wider, no restriction on the number of propagating modes being imposed. However, in the last case it should be mentioned that the differences between the currents carried by the *out*1 and *out*2 branches is larger as the number of propagating modes/the width of the interference region is smaller. The number of propagating modes in the interference region of width $W$ is given by the integer part of $(E-U_G)W/(\pi\hbar v_F)$.

## III. ROOM-TEMPERATURE MEASUREMENTS OF GRAPHENE-BASED QUANTUM LOGIC DEVICES

In this section we present experimental results related to the room-temperature operation of QG and DJ devices. We have measured all 50 QG and DJ devices using the Keithley 4200 SCS equipment with low noise amplifiers at outputs. The DC probes and probe station are embedded in a Faraday cage. The wafer containing QG and DJ devices is placed on the chuck of the probe station, the metallic terminals are connected by DC probes to the Keithley 4200 SCS and the Faraday cage is closed. The chuck is DC polarized to provide back-gate voltages to QG and DJ devices.

The QG and DJ devices were measured in the following way: a drain voltage was applied between the input *in* and a grounded output, say output *out*1, while the drain current was monitored at the other output, *out*2, at various backgate voltages. Then, the procedure is continued by interchanging the outputs, i.e. by a applying the drain voltage between *in* and the grounded *out*2 output and by measuring the current at *out*1. The gate currents were monitored during all measurements, and were found to be less than 1 pA, showing the effective action of the backgate voltage on the modulation of QG and DJ devices.

Due to the inherent differences between such small devices and/or imperfections in the fabrication step, not all quantum gates had identical electrical characteristics. However, similar characteristics are encountered in 45% of QG and DJ devices. The major source of the lack of reproducibility beyond 45% are grain boundaries defects in CVD grown graphene, which depreciate its physical properties; standard Raman analysis in different places of the 4 inch wafer have shown that about 78% of wafer is graphene monolayer. Another source of the lack of reproducibility beyond 45% are the unavoidable small misalignments, especially proximity effects, of the EBL process, which are important at such small dimensions. A higher repeatability could be achieved with a more advanced EBL equipment, with a higher resolution.

All measurements are performed at room temperature on all 50 quantum devices. The validity of our measurements is further tested by repeating the experiments at different voltage steps using dual-sweep procedures for measuring current-voltage dependences. No hysteretic behavior was observed and no changes in current-voltage dependences were detected. No smoothing procedures are used during the measurements or afterwards.

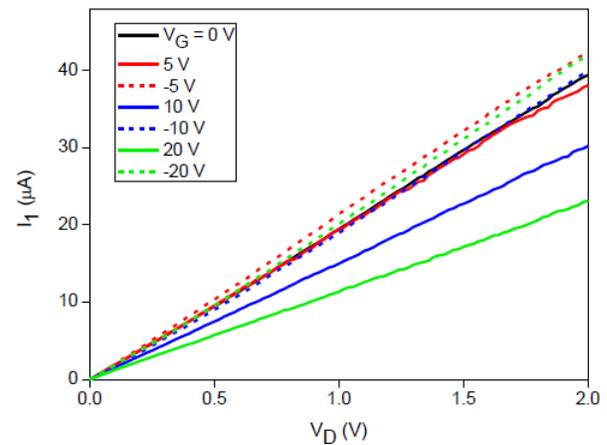

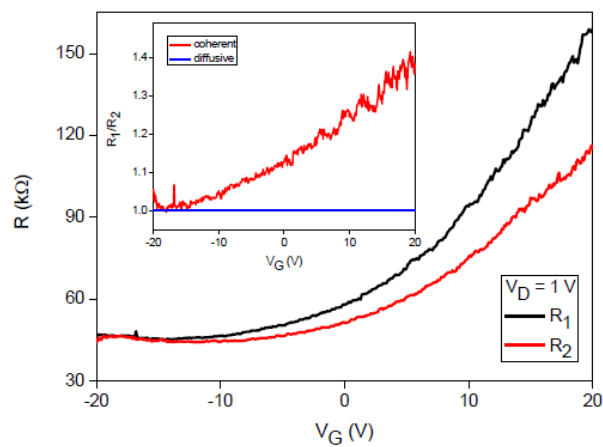

Fig. 2. Drain voltage dependence of the current at *out*1 in a QG device for different gate voltages in the legend. (b) Gate voltage dependence of the resistances at *out*1 and *out*2 for a drain voltage of 1 V in the same QG device. Inset: gate voltage dependence of the ratio of resistances at *out*1 and *out*2 as measured (red line) and as expected in the diffusive transport regime (blue line)

Figure 2(a) shows the linear drain voltage dependence of the current measured at *out*1, denoted as $I_1$, in a QG device for different gate voltage values indicated in the legend. From this figure it follows that the metallic pads make an ohmic contact with graphene, allowing an unimpeded current flow, and that the backgate modulates the current. Similar drain voltage dependences were obtained for the current $I_2$, measured at *out*2. Denoting as $R_1 = \partial V_G / \partial I_1$ and $R_2 = \partial V_G / \partial I_2$, the requirements $r1$ and $r2$ are satisfied if $R_1 \neq R_2$.

Indeed, this is the case, as can be seen from Fig. 2(b), which illustrates the gate voltage dependences of the resistances $R_1$, $R_2$ measured at *out*1 and *out*2 for a drain voltage of 1 V, as well as the $R_1/R_2$ ratio (red curve in the inset). At high positive $V_G$ values both resistances are high, which means that the device is *p*-doped during the fabrication process and that the number of propagating modes is small, but different at the two outputs. This indicates that the QG device is asymmetric. This asymmetry can be located at the input Y-junction, such that the coefficients of the superimposing $|0\rangle$ and $|1\rangle$ quantum logic states induced by it are different, or at the output Y-junction, i.e. at the splitting of the quantum wavefunction in the interference region into output wavefunctions, in *out*1 and *out*2. Either way, the dissimilarity in the emerging wavefunctions can occur in both amplitude and/or phase, satisfying condition $R_1 \neq R_2$.

As the gate voltage decreases, the resistances/currents at *out*1 and *out*2 ports become less dissimilar. Because in QG devices the output currents depend on the phase difference between the two propagating modes in the interference region of length $L$, given by $\Delta\phi(E, U_G, L) = L\left[\sqrt{(E-U_G)^2/(\hbar v_F)^2 - (2\pi/W)^2} - \sqrt{(E-U_G)^2/(\hbar v_F)^2 - (\pi/W)^2}\right]$, the ratio between *out*1 and *out*2 resistances/currents is influenced by the gate voltage. The dependence of the ratio between the two resistances on $V_G$ in Fig. 2(b) can then be attributed to the change in $\Delta\phi(E, U_G, L)$, which suggests that charge carrier propagation throughout the QG device is (quasi-)ballistic, i.e. coherent.

The fact that the $R_1/R_2$ depends on the gate voltage, as can be seen from the red curve in the inset of Fig. 2(b), proves that charge carrier is phase coherent, so that the device can be used for quantum computation. Indeed, a diffusive transport regime, in which the only influence of the gate voltage would be to modulate the concentration of carrier density, would lead to a ratio between the *out*1 and *out*2 resistances/currents dependent only on the dimensions of the output nanoribbons, and thus independent of $V_G$. To emphasize this fact, we have traced with blue line in the inset of Fig. 2(b) the expected dependence of the $R_1/R_2$ ratio on the gate voltage for diffusive transport. Although in this case the constant value of $R_1/R_2$ could have a value slightly different from 1, the essential point is that this ratio should not depend on $V_G$.

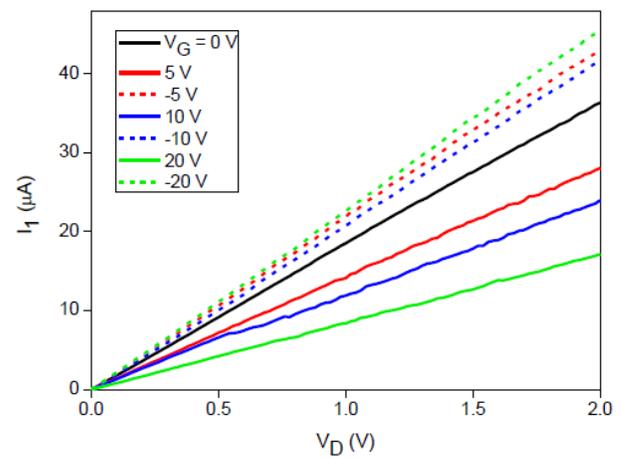

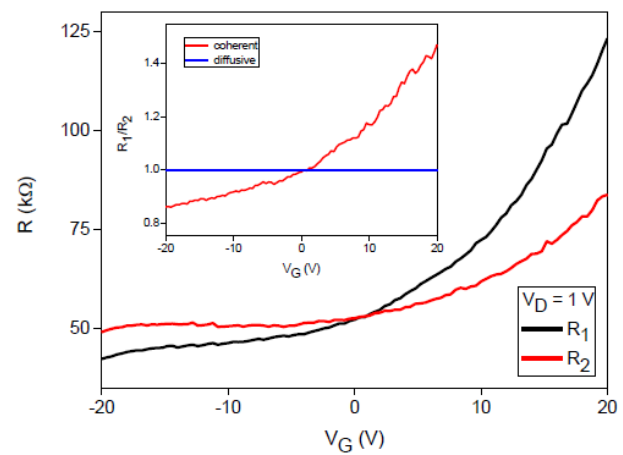

Fig. 3. Same as Fig. 2, for a different QG device.

Figures 3(a) and 3(b) illustrate similar dependences as Figs. 2(a) and 2(b), but in another QG device. Whether the general behavior of the two devices is comparable, in the second one the two resistances are almost the same at $V_G = 0$. It should be noted that, in both QG devices, the resistances generally decrease as the gate voltage decreases from positive to negative values (except for a very wide and almost imperceptible local maximum for *out*2 in the negative gate voltage region) due to another effect: the increase of the propagating mode number with a decrease in $V_G$. As this number increases, the eventual dissimilarity between the interfering quantum wavefunctions becomes less evident.

Both *p*-doped QG devices have not yet reached the Dirac point even at gate voltages as large as 20 V. As in Fig. 2(b), the measured $R_1/R_2$ ratio, plotted with red line in the inset of Fig. 3(b), shows a gate voltage dependence, which unambiguously demonstrates the coherent, i.e. (quasi-)ballistic, nature of charge carrier transport, in contrast to the expected result from a device with diffusive transport, in which the $R_1/R_2$ ratio would be independent of $V_G$, as suggested by the blue line in the same inset. Therefore, also this QG device could perform quantum operations.

On the other hand, DJ devices are designed to detect any asymmetric potential acting between the Y-junction and the interference region, which would affect the initial quantum superposition, $(|0\rangle+|1\rangle)/\sqrt{2}$, implemented by the symmetric Y-junction. In the absence of an asymmetric potential, the currents measured at ports *out*1 and *out*2 should be identical, any dissimilarity between these two currents indicating an asymmetry in the existing/applied potential. In this case, conditions *r*1 and *r*2 require that $R_1 = R_2$.

No asymmetric potential was applied during measurements of the fabricated DJ devices. Therefore, the *out*1 and *out*2 currents should be identical, irrespective of the backgate voltage, which has a constant (symmetric) value across the device. Contrary to QG devices, in this case requirement $R_1 = R_2$ implies that a voltage independent $R_1/R_2$ is not an indication of coherent transport, but, due to similar dimensions and technological procedures as those used for QG devices, we can assert that the fabricated DJ devices are (quasi-)ballistic.

Indeed, we have measured DJ devices with almost identical

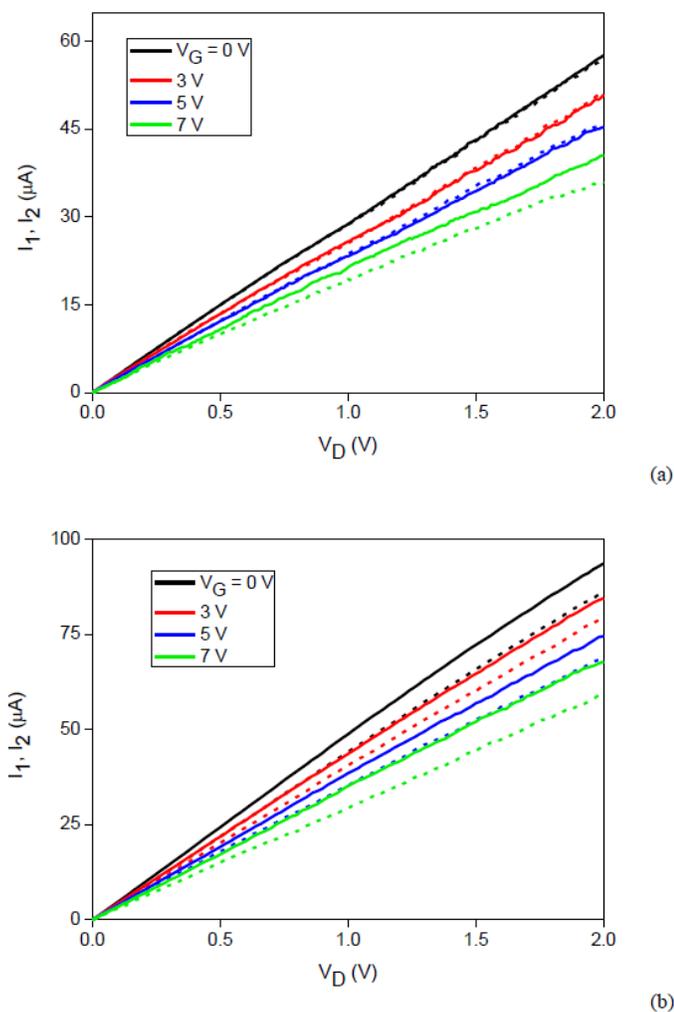

Fig. 4. Drain voltage dependence of the currents at *out*1 (solid lines) and *out*2 (dotted lines with the same color) ports for different gate voltages in the legend in the case of (a) a symmetric and (b) an asymmetric DJ device.

*out*1 (solid lines) and *out*2 (dotted lines with the same color) currents satisfying condition $R_1 = R_2$, at least for small gate voltages, as required. An example of such a situation is illustrated in Fig. 4(a). In this case, the two output currents are almost identical up to a gate voltage of 7 V, at which, probably because of a minor difference between the arms of either the input or output Y junction, the number of propagating modes starts to differ. In both DJ devices the contacts with the pads are ohmic.

However, we have also found that some DJ devices, as that in Fig. 4(b) have different *out*1 currents than *out*2 currents at the same drain voltage, for all applied gate voltages indicated in the legend. This means that the respective DJ devices are asymmetric themselves, more precisely that the splitting of the electron wavefunction at the Y-junctions is asymmetric. These asymmetries are due to technological imperfections and could be avoided by reducing the proximity effects of e-beam lithography.

Consequently, the functionalities as quantum gates of the devices presented in this paper were demonstrated using preliminary electrical measurements on 50 devices. The next step is the modulation of the initial quantum superposition in QG devices by in-plane electric fields that could be applied via adjacent gates in the splitting region of the input Y junction. This goal will not be easily achieved due the restrictions regarding alignments and minimum features required by e-beam lithography, but the work is in progress and will be the subject of a forthcoming publication.

## IV. CONCLUSIONS

We have fabricated at the wafer scale and electrically characterized graphene-based configurations that were predicted in previous papers to act either as one-qubit quantum gates or as one-qubit modified Deutsch-Jozsa algorithms. Our measurements confirmed the (quasi-)ballistic nature of charge carrier propagation through all devices. In the case of QG devices all measured structures showed dissimilarities at the input and/or output Y junctions, which diminish as the number of propagating modes increases, whereas in the case of DJ devices symmetric configurations have been identified. Both QG and DJ configurations were found, by preliminary measurements, to be suitable for quantum computation. However, much remains to be done to demonstrate a room-temperature, even rudimentary, graphene-based quantum computer: a full implementation of tunable one- and two-qubit quantum gates requires the possibility of modulating the initial quantum superpositions by in-plane electric fields and the fabrication of adjacent gates, and the sensitivity of DJ devices at detecting asymmetries in applied potentials remains to be tested. In addition, all quantum gates should be integrated with low-noise readout electronics.

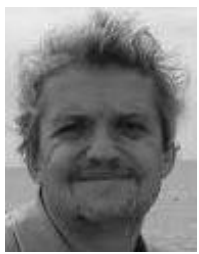

**Mircea Dragoman** has graduated the Electronic Faculty, Polytechnical Institute in Bucharest, in 1980, and received the doctoral degree in electronics in 1991. In the period 1992-1994 he was the recipient of the Humboldt Fellowship award and has followed postdoctoral studies at Duisburg University, Germany.

He is Senior Researcher I with the National Research Institute in Microtechnologies. He co-authored more than 250 scientific papers and 6 monographies, such as M. Dragoman, D. Dragoman, *Nanoelectronics. Principles and Devices*, at Artech House, Boston, USA (2008) and M. Dragoman and D. Dragoman, *2D Nanoelectronics*, *Physics and Devices of Atomically Thin Materials*, at Springer (2017).

Dr. Dragoman was awarded the Gheorghe Cartianu prize of the Romanian Academy in 1999.

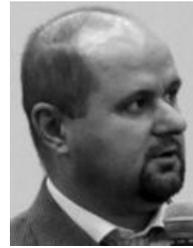

**Adrian Dinescu** received both M.Sc. degree (in 1993) and Ph.D. degree (in 2010) in solid state physics from the Faculty of Physics, University of Bucharest.

From 1993 to 1997, he was with the National Institute for Electronic Components Research, working in the field of optoelectronic components. Since 1997, he is with IMT-Bucharest, where at present he is the Head of Nanoscale Structuring and Characterization Laboratory. He is involved in micro- and nanoscale characterization using FE-SEM, and in structuring at the nanoscale using Electron Beam Lithography. His expertise also includes: micro- and nanofabrication, and optoelectronic measurements.

In the last 10 years, Dr. Dinescu has coauthored more than 100 papers in ISI ranked journals, and in 2016 was awarded with the Gheorghe Cartianu prize of the Romanian Academy.

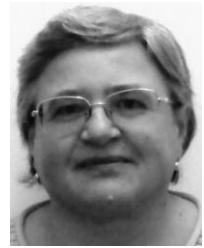

**Daniela Dragoman** received the M.Sc. degree in semiconductor physics from the Faculty of Physics, University of Bucharest, in 1989, and the Ph.D. degree in optoelectronics from the University of Limerick, Ireland, in 1993. Between 1998-199 and 2001-2002 she was the recipient of the Humboldt Fellowship award and has followed postdoctoral studies at University of Mannheim, Germany.

She is presently Professor with the Faculty of Physics, University of Bucharest. She has co-authored more than 280 scientific papers and 6 monographies, such as D. Dragoman and M. Dragoman, *Advanced Optoelectronic Devices*, at Springer (1999), D. Dragoman and M. Dragoman, *Quantum Classical Analogies*, at Springer (2004).

Dr. Dragoman was awarded the Gheorghe Cartianu prize of the Romanian Academy in 1999.